\documentclass[twocolumn,aps,prl,superscriptaddress,floatfix,longbibliography]{revtex4-1}  

\usepackage[draft]{changes}
\usepackage[normalem]{ulem}

\usepackage{graphicx}
\usepackage{amssymb} 
\usepackage{dcolumn}
\usepackage{bm}
\usepackage[mathlines]{lineno}
\usepackage[colorlinks,linkcolor=blue,anchorcolor=blue,citecolor=blue,urlcolor=blue]{hyperref}
\usepackage{multirow}
\usepackage{color}
\usepackage{amsmath}

\makeatletter
\renewcommand*{\@fnsymbol}[1]{\ensuremath{\ifcase#1\or \dagger\or *\or \ddagger\or
\mathsection\or \mathparagraph\or \|\or **\or \dagger\dagger \or
\ddagger\ddagger \else\@ctrerr\fi}} \makeatother

\usepackage{color}

\usepackage[draft]{changes}
\usepackage[normalem]{ulem}

\usepackage{color}%

\begin{document}
\title{
The role of high-order anharmonicity and off-diagonal terms in thermal conductivity: 
a case study of multi-phase CsPbBr$_3$
}
%

%

\author{Xiaoying Wang}
\affiliation{State Key Laboratory for Mechanical Behavior of Materials, Xi'an Jiaotong University, Xi'an 710049, China}

\author{Zhibin Gao}
\email[E-mail: ]{zhibin.gao@xjtu.edu.cn}
\affiliation{State Key Laboratory for Mechanical Behavior of Materials, Xi'an Jiaotong University, Xi'an 710049, China}

\author{Guimei Zhu}
\email[E-mail: ]{zhugm@sustech.edu.cn}
\affiliation{School of Microelectronics, Southern University of Science and Technology, 
Shenzhen, 518055, PR  China}



                
\author{Jie Ren}
\affiliation{Center for Phononics and Thermal Energy Science, China-EU Joint Center for Nanophononics, 
             Key Laboratory of Special Artificial Microstructure Materials and Technology, School of 
             and Engineering, Tongji University, Shanghai 200092, China}

\author{Lei Hu}
\affiliation{State Key Laboratory for Mechanical Behavior of Materials, Xi'an Jiaotong University, Xi'an 710049, China}
       
\author{Jun Sun}
\affiliation{State Key Laboratory for Mechanical Behavior of Materials, Xi'an Jiaotong University, Xi'an 710049, China} 
                
\author{Xiangdong Ding}
\affiliation{State Key Laboratory for Mechanical Behavior of Materials, Xi'an Jiaotong University, Xi'an 710049, China}

\author{Yi Xia}
\email[E-mail: ]{yxia@pdx.edu}
\affiliation{Department of Mechanical and Materials Engineering, Portland State University, Portland, Oregon 97201, USA}



\author{Baowen Li}
\affiliation{Department of Materials Science and Engineering, Department of Physics, Southern University of Science 
                and Technology, Shenzhen, 518055, PR China. Paul M. Rady Department of Mechanical Engineering and Department of Physics, University of 
                Colorado, Boulder, Colorado 80305-0427, USA}   

\date{\today}
\begin{abstract}


We investigate the influence of three- and four-phonon scattering, perturbative anharmonic 
phonon renormalization, and off-diagonal terms of coherent phonons on the thermal conductivity of CsPbBr$_3$ phase change perovskite, by using 
advanced implementations and first-principles simulations. Our study spans a wide temperature range covering the entire structural spectrum.
%
%
%
%
%
Notably, we demonstrate that the interactions between acoustic and optical phonons 
result in contrasting trends of phonon frequency shifts for the high-lying
optical phonons in orthorhombic and cubic CsPbBr$_3$ as temperature varies. 
%
%
%
Our findings highlight the significance of wave-like tunneling of coherent phonons 
in ultralow and glass-like thermal conductivity in halide perovskites.
%
%
%
%
%
\end{abstract}



\maketitle



%

\section{I. Introduction}
Phonons are crucial for understanding thermal transport in semiconductors and dielectrics. 
In the traditional approach of first-principles calculations, important quantities such as the mode-Grüneisen 
parameter, thermal expansion, phonon group velocity, three-phonon lifetime, and linewidth can 
be obtained by
using anharmonic lattice dynamics 
under the quasi-harmonic approximation through the linearized phonon Boltzmann 
transport equation~\cite{phono3py, carrete2014finding, van2016high}. 


However, the traditional approach, in which some important factors have been overlooked, is facing several challenges:

1) the higher-order interatomic interactions like the fourth-order have been ignored for a long time like in BAs 
in which the four-phonon scattering is responsible for around 40\% suppression of $\kappa_L$ compared to that with only
three-phonon interactions;

2) interatomic anharmonicity increases as temperature is raised. However, the 
perturbation approach is difficult to deal with highly anharmonic systems, such as cubic ABX$_3$ perovskites 
with imaginary frequencies of harmonic phonons~\cite{tadano2015self, zhao2021lattice}. Therefore, we should 
illustrate the significance of anharmonic phonon renormalization, by utilizing the self-consistent phonon (SCPH) with temperature-dependent frequencies in several systems~\cite{xia2020high, kang2019intrinsic};

3) the off-diagonal terms in the heat flux operator representing the heat transfer through the tunneling of wave-like coherent phonons could provide a potential method 
to bridge the gap between the traditional Peierls-Boltzmann transport equation and the experimental measurement in 
ultralow thermal conductivity materials and glass-like materials~\cite{simoncelli2019unified}. Items 1) and 2) are 
related but independent from 3). %

CsPbBr$_3$ is a classical chalcogenide material and a promising candidate for thermoelectric application,
garnering significant attention in recent years. %
However, current studies have mainly focused on the thermal transport properties 
of individual phase~\cite{simoncelli2019unified, tadano2022first}, often considering only 
three-phonon scattering and calculating the lattice thermal conductivity with zero temperature 
phonon dispersion~\cite{hu2021metal,lee2017ultralow}.

The systematic investigation of all three 
phases of CsPbBr$_3$, considering both renormalization effects and four-phonon scattering, as well 
as the contribution of off-diagonal terms, remains largely unexplored. Previous studies have 
primarily focused on phase transitions, vibrational mechanisms, and dielectric 
properties~\cite{lanigan2021two, svirskas2020phase}. %
To the best of our knowledge, no systematic studies have been conducted to date that 
encompass all three phases while considering renormalization, four-phonon scattering, and 
off-diagonal contributions. %



In this study, we conduct a systematic investigation of the influence of quartic anharmonicity 
on the lattice dynamics and thermal transport properties of the three distinct phases of 
CsPbBr$_3$. We employ recent advancements in first-principles simulations, incorporating:
(i) efficient construction of high-order interatomic force constants (IFCs) from the CSLD 
method~\cite{zhou2014lattice, zhou2019compressive, zhou2019compressive2}; (ii) rigorous 
calculations of temperature-dependent phonons through SCPH theory and higher-order multi-phonon 
scattering rates~\cite{tadano2015self, feng2016quantum}; (iii) evaluation of the lattice thermal 
conductivity $\kappa_L$ by using a unified theory that considers both diagonal terms from the 
standard Peierls contribution and off-diagonal terms from the coherent Wigner 
distribution~\cite{simoncelli2019unified, isaeva2019modeling}. %

\section{II. COMPUTATIONAL METHODS}
For the Peierls-Boltzmann transport equation, the lattice thermal conductivity $\kappa_p$ can be calculated as
%
\begin{eqnarray}
\label{eqn1}
\kappa_p=\frac{\hbar^2}{k_B T^2 V N_0} \sum_{\lambda} n_\lambda (n_\lambda + 1) \omega^2_\lambda \bm{v}_\lambda \otimes \bm{v}_\lambda \bm{\tau}_\lambda,
\end{eqnarray}
where $\hbar$, $k_B$, $T$, $V$, $N_0$ are the reduced Planck constant,
Boltzmann constant, absolute temperature, primitive unit cell volume, and 
the total number of sampled phonon wave vectors in the first Brillouin zone, respectively. %
$n_\lambda$, $\omega_\lambda$,  $v_\lambda$, and $\tau_\lambda$
are the equilibrium component
of the phonon population
, frequency, group velocity, and lifetime for the 
$\lambda$ mode (wave vector $q$ and  branch index $s$), respectively. %
Except for $\tau_\lambda$, all the above parameters can be 
obtained from harmonic approximation (HA). Usually, $\tau_\lambda$ 
can be obtained from the perturbation 
theory by consideration of three-phonon scattering~\cite{debernardi1995anharmonic, gao2018unusually}. %

\begin{figure*}
\includegraphics[width=2.0\columnwidth]{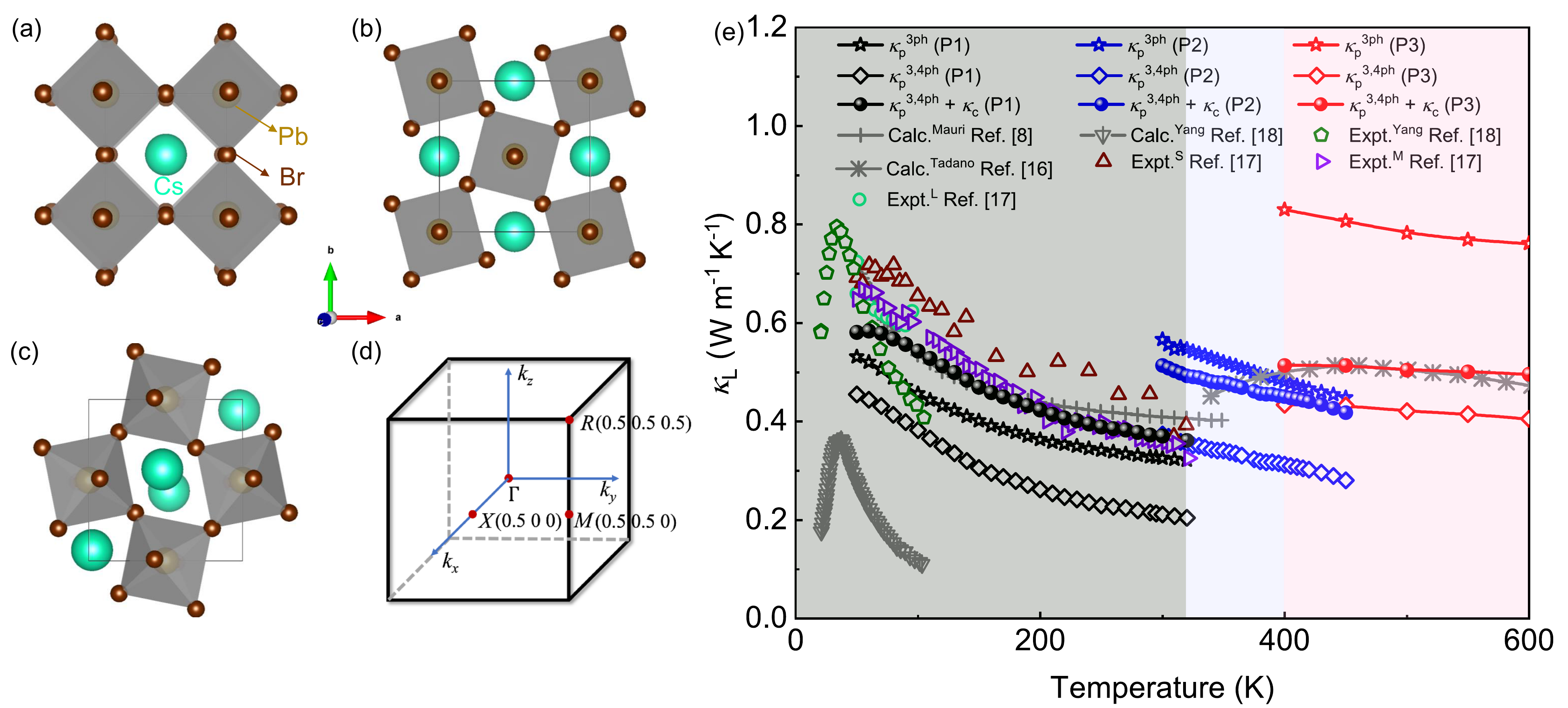}
\caption{%
(a)--(c) Crystal structures of CsPbBr$_3$ in the orthorhombic, tetragonal, and cubic 
phases, respectively. The green, brown, and yellow colors represent Cesium (Cs), %
Bromine (Br), and lead (Pb) atoms.  (d) The First Brillouin zone of the cubic phase with 
high symmetry points $\Gamma$, $X$, $M$, and $R$ indicated by red dots.  (e)
Lattice thermal conductivity $\kappa_L$ of CsPbBr$_3$ includes 
diagonal and off-diagonal contributions in three phases using self-consistent phonon approximation. %
P1, P2, and P3 represent orthorhombic, %
tetragonal, and cubic phases. $\kappa_p$ (Eq. (1)) is the standard Peierls 
contribution, and $\kappa_c$ (Eq. (4)) is the coherent contribution from the 
off-diagonal Wigner distribution elements. 3ph indicates only three-phonon
scattering is included, and 3, 4ph means both three-phonon and four-phonon scattering are considered. For 
comparison, the unified first-principles theory~\cite{simoncelli2019unified} 
and the quasiparticle nonlinear theory (QP-NL)~\cite{tadano2022first} are plotted for reference. %
Expt.$^L$, Expt.$^M$, and Expt.$^S$ refer to experiments on single crystalline nanowires~\cite{wang2018cation} with cross
sections of 800 $\times$ 380 nm$^2$, 320 $\times$ 390 nm$^2$, and 300 $\times$ 160 nm$^2$, respectively. %
%
\label{fig1}}
\end{figure*}


The temperature-dependent phonon dispersion could be 
considered by the anharmonic phonon renormalization (APRN) at  finite temperatures~\cite{li2010energy, souvatzis2008entropy, errea2011anharmonic, errea2014anharmonic, xia2018revisiting, ravichandran2020phonon}. 
%
Among various existing approaches, SCPH \cite{tadano2015self,xia2020high} approximation is one effective method that can rigorously account for the first-order correction of phonon frequencies from the quartic anharmonicity. It can better describe the soft phonon modes and strong anharmonicity. In brief, under the SCPH approximation, the 
temperature-dependent renormalized phonon frequency  $\Omega_\lambda$ can be obtained from the following equation~
\begin{eqnarray}
\label{eqn2}	 
{\Omega}_{\lambda}^{2} = {\omega}_{\lambda}^2+2{\Omega}_{\lambda}\sum\limits_{{\lambda}_{1}} I_{\lambda\lambda_1},
\end{eqnarray}
where $\omega_\lambda$ is the original phonon frequency from 
the harmonic approximation. The scalar 
$I_{\lambda\lambda_1}$ can be obtained as,
\begin{eqnarray}
\label{eqn3}	 
{I_{\lambda\lambda_1}}=\frac{\hbar}{8 N_0} \frac{V^{(4)} (\lambda,-\lambda,\lambda_1,-\lambda_1)}{\Omega_{\lambda}\Omega_{\lambda_1}} \left[1+2n_\lambda(\Omega_{\lambda_1})\right],
\end{eqnarray}
in which $V^{(4)}$ is the fourth-order IFCs in the reciprocal representation. The phonon population $n_\lambda$ satisfies Bose-Einstein distribution as a function of temperature. Both Eq. (\ref{eqn2}) and Eq. (\ref{eqn3}) have parameters $I_{\lambda\lambda_1}$ and $\Omega_\lambda$ in common, and thus the SCPH equation can be solved iteratively. 
Note that $I_{\lambda\lambda_1}$ can be interpreted as the interaction between a pair of phonon modes, $\lambda$ and $\lambda_1$ including the temperature effects~\cite{tadano2015self, xia2020high}. %

\begin{figure*}
\includegraphics[width=2.0\columnwidth]{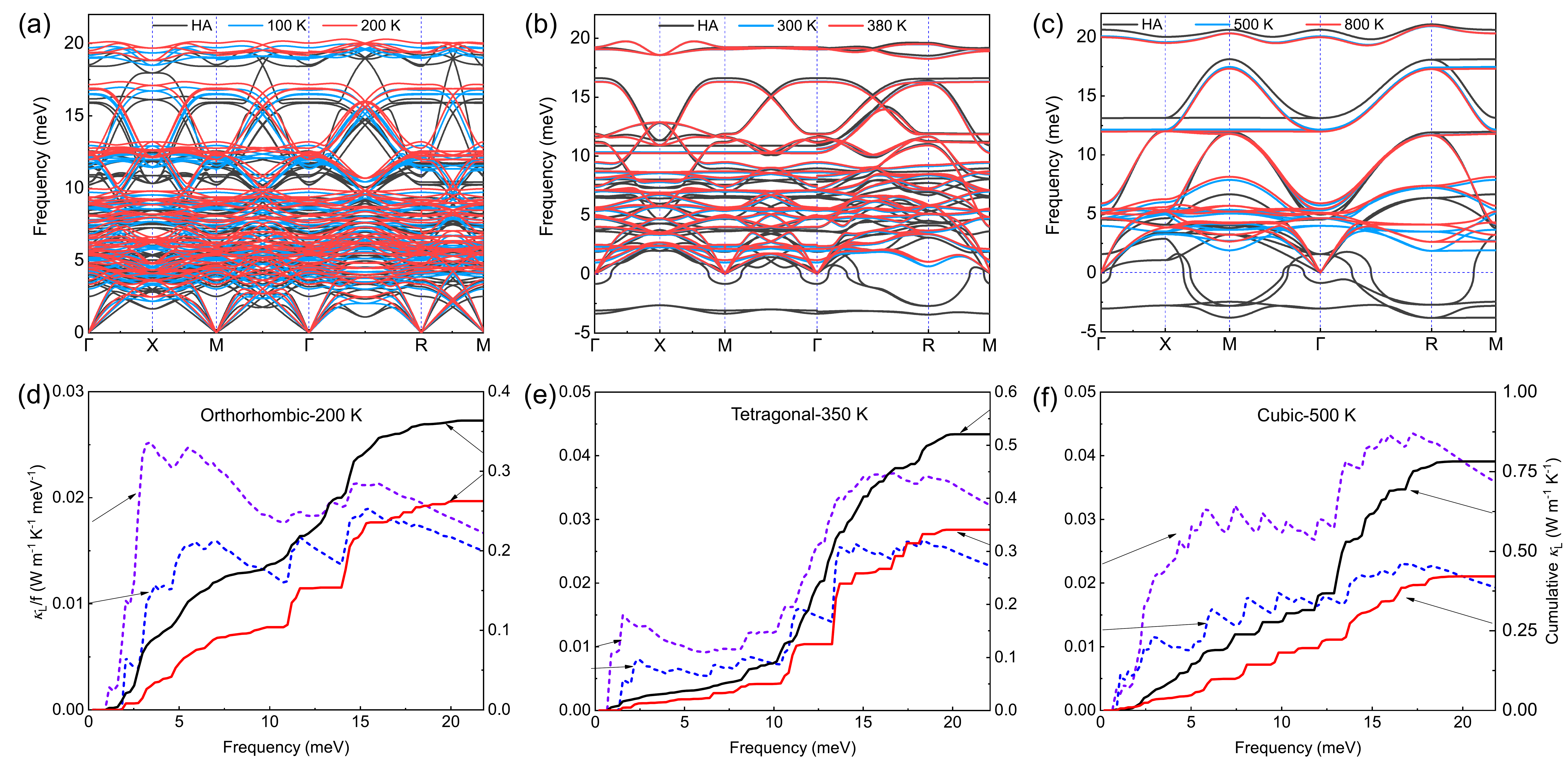}
\caption{%
Renormalized phonon dispersions for (a) orthorhombic, (b) tetragonal, and (c) cubic phases at different temperatures, respectively. HA is the harmonic approximation.
~(d), (e), and (f) are the frequency-resolved $\kappa_L$ (dashed lines) and cumulative $\kappa_L$ (solid lines) using 3ph (the upper line) and 3,4ph 
~(the lower line) methods at different temperatures, respectively.
\label{fig2}}
\end{figure*}

Moreover, if one considers the off-diagonal terms of the heat-flux operator, which depicts the tunneling of coherent phonons, an additional contribution of lattice thermal conductivity, $\kappa_c$, needs to be considered~\cite{semwal1972thermal, knauss1974theory, srivastava1981diagonal}. Usually, $\kappa_c$ is neglected in simple crystals because of well-separated phonon dispersions and slight broadening as a function of temperature. However, it could dominate in disordered and glass-like amorphous compounds where phonon and related group velocities cannot be clearly defined, and heat transfer is mediated by diffusons and locons~\cite{allen1989thermal, allen1993thermal, cepellotti2016thermal}. %

Recent studies show that $\kappa_c$ is substantial for materials with ultralow thermal conductivity, such as Mn$_4$Si$_7$ with twisting phonons~\cite{chen2015twisting}, Ba$_{7. 81}$Ge$_{40. 67}$Au$_{5. 33}$ clathrate~\cite{lory2017direct}, and Tl$_3$VSe$_4$~\cite{mukhopadhyay2018two}. %
Therefore, in all-inorganic halide perovskite CsPbBr$_3$, we incorporate $\kappa_c$ as follows, %
\begin{equation}
\begin{split}
\label{eqn4}	 
\kappa_c  =& \frac{\hbar^2}{k_B T^2 V N_0} \sum_{\bm{q}} \sum_{s \neq s'} \frac{ \omega^s_{\bm{q}} + \omega^{s'}_{\bm{q}} }{2}  \bm{v_q^{s, s'} \otimes v_q^{s', s}}\\
& \times \frac{ \omega^s_{\bm{q}}  n^s_{\bm{q}} ( n^s_{\bm{q}} + 1 ) + \omega^{s'}_{\bm{q}} n^{s'}_{\bm{q}} ( n^{s'}_{\bm{q}}  +1 )  }{ 4 ( \omega^{s'}_{\bm{q}} - \omega^s_{\bm{q}}  )^2  + ( \Gamma^s_{\bm{q}} + \Gamma^{s'}_{\bm{q}} )^2 } ( \Gamma^s_{\bm{q}} + \Gamma^{s'}_{\bm{q}} ),
\end{split}
\end{equation}
where the phonon lifetime in Eq. (\ref{eqn1}) is substituted as 
$\Gamma^s_{\bm{q}} = 1/ \tau_\lambda$, including three-phonon (3ph) and four-phonon (4ph) scattering.
~The group velocity is replaced with a generalized form containing off-diagonal 
elements~\cite{simoncelli2019unified, allen1993thermal},
\begin{equation}
\label{eqn5}	 
\bm{{v_q}} ^{s', s} = \frac{ \left \langle \bm{e_q} ^s  \right |  \frac{ \partial D ( \bm{{q}} ) }{ \partial \bm{{q}} }  | \bm{e_q} ^{s'}  \rangle   }{  \omega^{s}_{\bm{q}} + \omega^{s'}_{\bm{q}} },
\end{equation}
in which $\bm{e_q} ^s$ and $D( \bm{{q}} )$ are the polarization vector and the phonon dynamical matrix, respectively. When $s=s'$, it stands for the phonon band diagonal terms, while $s \neq s'$, corresponds to the off-diagonal terms. 

Therefore, the total lattice thermal conductivity $\kappa_L = \kappa_p + \kappa_c$. Note that in order to compute the generalized group velocity correctly, we used the phase convention that accounts for atomic positions within its lattice point to construct the dynamical matrix, as adopted in earlier studies \cite{PhysRevB.29.2884}. The details of the calculation are shown within the 
Supplemental Material~\cite{supp, kresse1996efficient, blochl1994projector, kresse1999ultrasoft, esfarjani2008method, togo2015first,candes2008introduction}.

%

 %



\section{III. RESULTS AND DISCUSSION}
CsPbBr$_3$ is a typical phase change material of ABX$_3$ perovskite. 
One can discriminate the transition temperature by the dynamical instability of the appearance of soft acoustic phonons from anharmonic potential energy surfaces~\cite{he2020anharmonic, krapivin2022ultrafast}. %
%
Specifically for CsPbBr$_3$, a second-order phase transition occurs at about 318 K, and the
transition temperature may vary, up to 361 K, depending on different samples. %
At about 373 K, there is another first-order phase transition. The critical temperature was also found at a higher temperature of about 401 K due to different experimental conditions~\cite{natarajan1971phase,malyshkin2020new}.
Here, we have chosen an intermediate temperature by considering the above different experimental values~\cite{natarajan1971phase,malyshkin2020new}. Accordingly, the estimated temperatures for the first-order and second-order phase transitions are 320 K and 400 K, respectively.
%
%
%
The high-temperature phase remains a cubic symmetry. The temperature reduction induces symmetry-breaking, leading to anisotropic structures from the cubic to the tetragonal at mid-temperature and finally to the orthorhombic crystals at low temperature~\cite{lanigan2021two, bechtel2019finite, malyshkin2020new, natarajan1971phase}. All three crystal structures are depicted in Fig.~\ref{fig1} (a-c). %
%


Fig.~\ref{fig1}(e) shows the effects of SCPH, 4ph, and $\kappa_c$ on the 
calculated lattice thermal conductivity
of CsPbBr$_3$.
In the following, we neglect the SCPH notation for simplicity. Different 
primitive cells are used to calculate the corresponding temperature 
range. %
Due to the different crystal symmetry, we find that $\kappa_L$ increases from the orthorhombic to the tetragonal and the cubic phases. %
Moreover, $\kappa_p^{3,4ph}$+$\kappa_c$ in each phase decreases as temperature increases because of the enhanced phonon scattering. %
Since P1 and P2 are anisotropic, we use the arithmetic mean value in the figure. 

Compared with 
$\kappa_p^{3ph}$, $\kappa_p^{3,4ph}$ is smaller due to the additional 4ph 
scattering. Moreover, the gap between them ($\Delta =  \kappa_p^{3ph} - \kappa_p^{3,4ph}$) 
is growing significantly from P1 to P2 and finally to P3 based on Eq. (\ref{eqn1}). %
For instance, $\Delta$ is 0.114, 0.168, %
and 0.361 W m$^{-1}$ K$^{-1}$
for temperature 300 K, 400 K, and 500 K,
respectively. %
It is usually attributed to the different scaling laws of 4ph ($\tau_4^{-1} \sim T^{2}\omega^{4}$) 
and 3ph ($\tau_3^{-1} \sim T\omega^{2}$) scatterings in which $\tau$ is the relaxation time~\cite{feng2017four}. %
Therefore, 4ph scattering is more critical than 3ph scattering at high-temperature 
and $\Delta$ is proportional to the temperature. %

%

Since the ABX$_3$
~perovskite has ultralow $\kappa_L$ and off-diagonal terms contribute 
significantly~\cite{lee2017ultralow, simoncelli2019unified}, we include $\kappa_c$ 
calculation of CsPbBr$_3$ based on Eq. (\ref{eqn4}) and Eq. (\ref{eqn5}). At 300 K, the value of 
$\kappa_c$ of CsPbBr$_3$ for P1 phase is 0.158 W m$^{-1}$ K$^{-1}$. 
At 400 K and 500 K, $\kappa_c$
of CsPbBr$_3$ for both P2 and P3 phases are 0.138 and 0.084 W m$^{-1}$ K$^{-1}$, 
respectively. More details
of the values can be found in Supplemental Material S3~\cite{supp}. %
%
%

Our results of $\kappa_c$ agree reasonably well with the one reported by Simoncelli, Marzari, and Mauri~\cite{simoncelli2019unified}, while the minor deviations might come from the size of the supercell used in the calculation of harmonic phonon as well as additional effects arising from quartic anharmonicity. %
%
%
For the P3 phase, the value 
of ($\kappa_p^{3,4ph}+\kappa_c$) is 0.501 W m$^{-1}$ K$^{-1}$ 
at 500 K, %
which is quite close to the result of 0.50 W m$^{-1}$ K$^{-1}$ from the recent quasiparticle nonlinear theory (QP-NL)~\cite{tadano2022first}. %

\begin{figure}
\includegraphics[width=1.0\columnwidth]{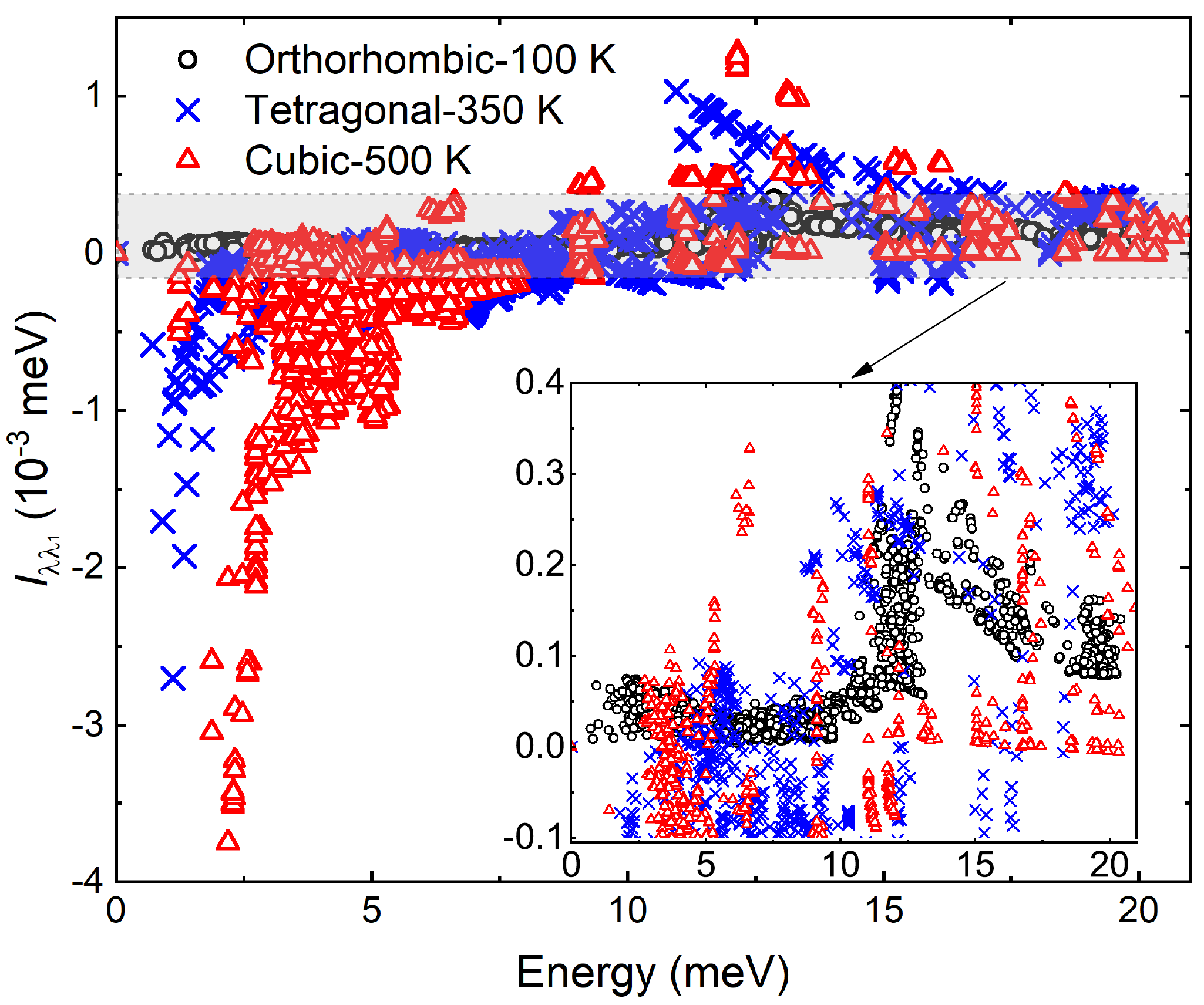}
\caption{
Scattering strength $I_{\lambda\lambda_1}$ of the 4ph interaction matrix elements 
between the highest zone-center optical phonon mode and all other remaining phonons
in three various phases according to Eq. (\ref{eqn2}) and Eq. (\ref{eqn3}) for the (a) 
orthorhombic at 100 K, (b) tetragonal at 350 K, and (c) cubic phase at 500 K, respectively. 
%
\label{fig3}}
\end{figure}

%
%

%
%
%

Next, we investigate the influence of anharmonic renormalization
on  phonon dispersion among orthorhombic, tetragonal, and cubic phases of CsPbBr$_3$. %
Phonon-phonon interaction and lattice anharmonicity are ascribable to the cubic, %
quartic, and even higher-order IFCs. 

The phonon spectra
at different temperatures in three phases are shown in Fig.~\ref{fig2}. Unexpectedly, %
we notice that the acoustic and optical phonon branches become hardened 
as temperature increases for the low-temperature orthorhombic phase of CsPbBr$_3$, shown in
Fig.~\ref{fig2}(a). %
Nevertheless, it is indisputable for the high-temperature cubic phase that the 
acoustic branches stiffen, whereas the top three optical branches soften with 
increasing temperature, shown in Fig.~\ref{fig2}(c). %
For the tetragonal phase of CsPbBr$_3$ in Fig.~\ref{fig2}(b), the high-lying
optical phonons is almost temperature independent.
We also plot the off-diagonal term contribution for CsPbBr$_3$ heat transport of three 
phases, which can be found in the Supplemental Material~\cite{supp}. %

%

We also analyze the frequency-resolved (dashed lines) and cumulative (solid lines)
lattice thermal conductivity $\kappa_L$ at different temperatures for cubic CsPbBr$_3$, shown in 
Fig.~\ref{fig2}(d-f). %
Since 4ph has proved to be additional scattering, $\kappa_L$ of 3ph (the 
upper one) is more significant than that of 4ph (the lower one). %

Traditionally, acoustic phonons are the main heat carriers.
However, 
in cubic CsPbBr$_3$, it is found that optical phonons, ranging from 13.0 to 16.0 meV, %
dominate the heat transport, no matter whether or not the 4ph scatterings processes are accounted for. %
Furthermore, 4ph scattering reduces $\kappa_L$ of the cubic CsPbBr$_3$ by almost 40$\%$ on top of 3ph. %


It is noticed that phonons of various frequencies dominate $\kappa_L$ among different phases. Optical phonons above the frequency of 15.0 meV control the heat transport of 3ph and 4ph for the orthorhombic phase. In comparison, acoustic and optical phonons among the frequency of 3.0 meV and 6.0 meV are also important for 3ph transport. However, for the tetragonal phase, we can find that optical phonons among the frequency from 13.0 meV to 19.0 meV dominate the phonon transport of 3ph and 4ph as well. 

Different phonon-temperature tendencies is an interesting phenomenon that has yet to be thoroughly investigated.
%
%
Previous work only found consistently softened or hardened
phonons~\cite{zheng2022anharmonicity, zhao2021lattice} in different materials. %
For the first time, we unveiled the multi-tendency of high-frequency phonon 
modes variation occurring in different phases of the same material. We reveal in the following that such an opposite tendencies of optical phonon modes of cubic CsPbBr$_3$ 
as a function of temperature is sourced from the interaction between 
the top three opticals and other phonon modes. %

\begin{figure*}
\includegraphics[width=2.0\columnwidth]{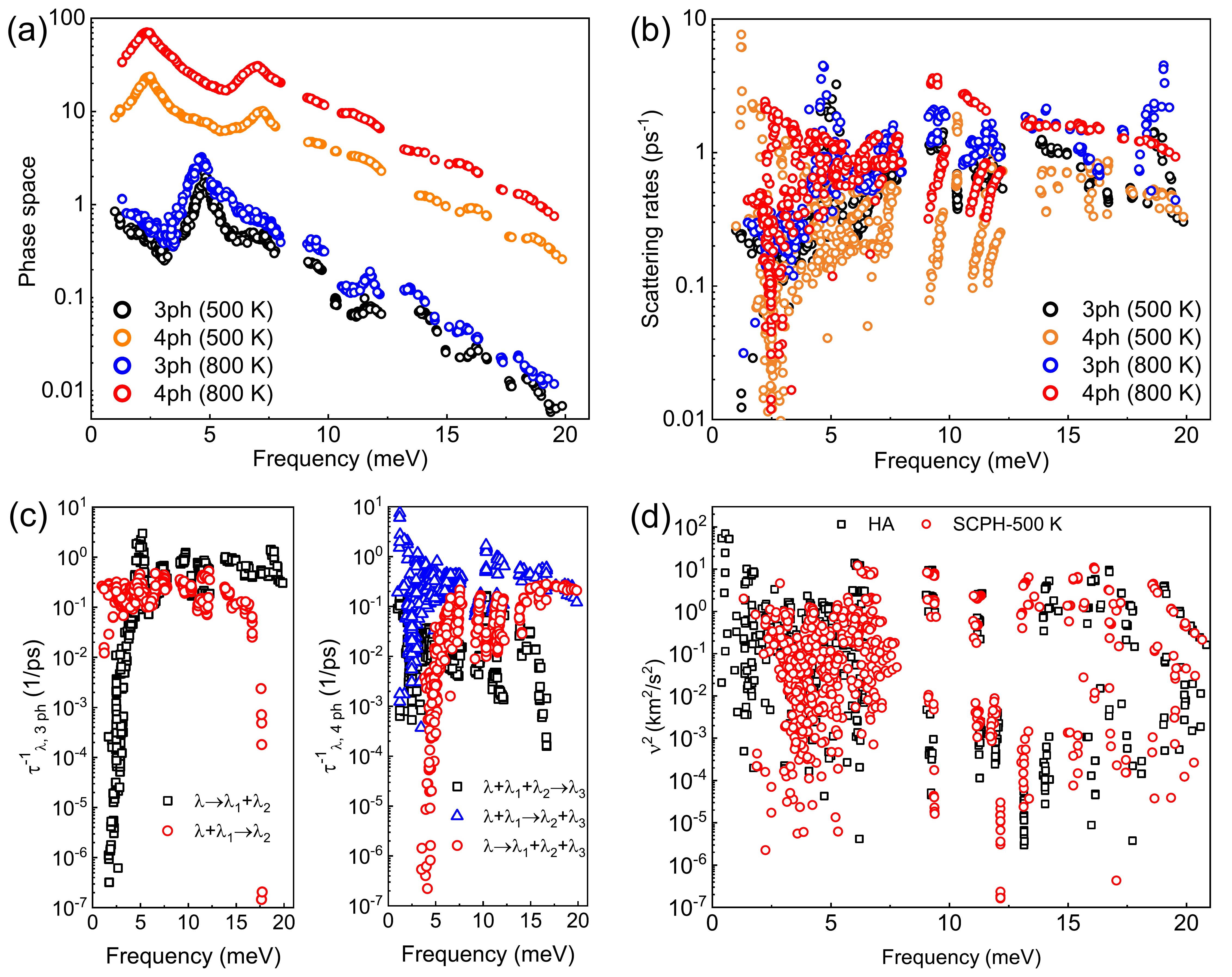}
\caption{
(a) Phonon scattering phase space and (b) scattering rates between 3ph and 4ph for 
cubic CsPbBr$_3$ at different temperatures. (c) 3ph and 4ph scattering 
with channels resolution, including the splitting ($\lambda {\to} \lambda_1 + \lambda_2$, 
$\lambda {\to} \lambda_1 + \lambda_2 + \lambda_3$), combination ($\lambda 
+ \lambda_1 {\to}  \lambda_2$, $\lambda + \lambda_1 + \lambda_2 {\to}  \lambda_3$),
and redistribution ($\lambda + \lambda_1 {\to} \lambda_2 + \lambda_3$) processes
at 500 K. (d)  $v^2$, where $v$ is Phonon group velocity. %
%
%
\label{fig4}}
\end{figure*}


In order to understand the underlying physical mechanism of the optical branches with high frequencies of 
different CsPbBr$_3$ phases have opposite temperature dependence in
their phonon spectra, %
%
we have systematically studied the strength of 4ph interaction matrix elements
$I_{\lambda\lambda_1}$ that have been introduced in Eq. (\ref{eqn2}) and Eq. (\ref{eqn3}). %
%
%
We set $\lambda$ for the highest optical phonon mode as any mode of the three highest optical phonon branches (index of mode=13-15 for cubic 
and 58-60 for the orthorhombic phase) %
and change $\lambda_1$ from the lowest acoustic 
phonon (index of mode =1) to all other optical phonons, 
gradually scrutinizing the    
interaction between phonon population $n_\lambda$, quartic-anharmonicity 
$V^{(4)} (\lambda,-\lambda,\lambda_1,-\lambda_1)$, phonon frequencies 
$\Omega_{\lambda_1}$, and $\Omega_{\lambda}$, respectively. %
%
%
Here we use the imode parameter to label the index of the phonon branches. %
%

Interestingly, $I_{\lambda\lambda_1}$ is mainly positive for the orthorhombic phase,  as is shown in Fig.~\ref{fig3}. It can be both positive and negative for the tetragonal phase, and they almost cancel with each other leading to a small net frequency change.
In contrast, it is found that the highest three optical branches 
have strong coupling with phonons 
in the low-frequency region, and the interaction $I_{\lambda\lambda_1}$ is even 
negative in the cubic CsPbBr$_3$, shown in Fig.~\ref{fig3}. We further examine 
the wave vector position behind the negative $I_{\lambda\lambda_1}$ and discover 
that most negative $I_{\lambda\lambda_1}$ stem from the low-energy acoustic phonon 
modes, especially around $M$ and $R$ high-symmetry points. %
%
%

In Supplemental Material S9-S11~\cite{supp}, we also show the $I_{\lambda\lambda_1}$ between imode=3, 57 at 100 K 
for the P1 phase, imode=3, 27 at 350 K for the P2 phase, and imode=3, 12 at 
500 K for the P3 phase, respectively. Since $I_{\lambda\lambda_1}$ can be either positive or negative, the renormalized phonon frequency $\Omega_\lambda$ 
as a function of temperature can either increase or decrease according to
Eq. (\ref{eqn2}). 
Based on Eq. (\ref{eqn3}), only when $V^{(4)} (\lambda,-\lambda,\lambda_1,-\lambda_1)$ is negative
for the three highest optical phonons of cubic phase, leading to a negative $I_{\lambda\lambda_1}$ 
and a reduced renormalization phonon frequency based on Eq. (\ref{eqn2}). %
On the contrary, $I_{\lambda\lambda_1}$ is positive 
as a function of temperature for the orthorhombic phase, and finally results 
in an increased $\Omega_\lambda$. Owing to the strong interaction between 
low-energy phonon modes around $M$ and $R$ and high-frequency optical
phonons, the three highest optical phonon frequencies, with the temperature increasing,
are softening for the cubic phase while hardening for the orthorhombic phase. 
More details can be found in the Supplemental Material~\cite{supp}. %
%

To further understand the effects of anharmonic phonon renormalization and 4ph scattering
on the thermal transport properties of CsPbBr$_3$, we continue to examine several parameters related to the
lattice thermal conductivity, i.e., phonon phase space and scattering rates, respectively.

All 
available 3ph and 4ph scattering phase spaces 
 need to satisfy the energy 
and quasi-momentum conservation simultaneously~\cite{lindsay2008three, gao2018unusually}, shown in Fig.~\ref{fig4}(a). %
The phase space of 3ph and 4ph scatterings increases as the temperature rises from 500 K to 800 K. Since the unit of 
phase space of 3ph and 4ph is different, one cannot compare them directly. %
%
%
Nevertheless, the larger phase space means more available scattering channels. %
The scattering strength in each accessible channel determines the final phonon relaxation time. %
%
%
%
Therefore, by including 4ph scattering, the lattice thermal conductivity is generally smaller than that 
of with only 3ph scattering. %
%

The phonon scattering rates of the cubic phase are shown in Fig.~\ref{fig4}(b). It displays that 4ph scattering has the same order of phonon scattering strength as that of 3ph. %
The scattering rates and phase space results for the orthorhombic and the tetragonal structures also show the same trends in 
Supplemental Material S5-S8~\cite{supp}.

Fig.~\ref{fig4}(c) displays the absorption and emission processes of the 3ph 
and 4ph as a function of frequency at 500 K, respectively. For the 3ph scattering, we consider the phonon 
splitting ($\lambda {\to} \lambda_1 + \lambda_2$) and combination ($\lambda 
+ \lambda_1 {\to}  \lambda_2$). For the 4ph situation, we count both phonon splitting ($\lambda {\to} \lambda_1 + \lambda_2 + \lambda_3$), %
and combination ($\lambda + \lambda_1 + \lambda_2 {\to}  \lambda_3$), as well as
redistribution ($\lambda + \lambda_1 {\to} \lambda_2 + \lambda_3$) processes. %
In the low-frequency region wherein acoustic modes dominate, 3ph combination processes are stronger than 
the splitting situation, while the redistribution processes of 4ph are the dominant 
ones. However, in the high-frequency region that is dominated by optical modes, the splitting process of 3ph 
becomes more important. For 4ph scattering, the splitting process also increases to a dominating portion and has the same order as the redistribution process.

Fig.~\ref{fig4}(d) shows the temperature effect on $v^2$, where $v$ is the group velocity
at different temperatures for the cubic phase. Interestingly, $v^2$ of 800 K is higher 
than that of 500 K for most frequencies %
but almost the same for the acoustic phonons. %
%
It stems from the renormalized phonon dispersions at finite temperatures. %
Besides, we distinguish phonon group velocity for the cases with and without considering SCPH in Supplemental Material S4~\cite{supp}.  

Besides, for the ABX$_3$ perovskites, we find that dynamical 
stability is in line with the thermodynamic stability~\cite{yang2022atlas,sun2017thermodynamic}. %
Previous work also used finite-temperature phonon dispersion of different phases of perovskites 
to predict the phase transition temperature~\cite{tadano2022first}. But for other materials, dynamical 
stability and thermodynamical stability have no direct connection. %

In general, the higher the lattice constant, the weaker the interatomic 
interaction in materials, which will usually lead to a lower lattice thermal conductivity. %
On the one hand, the calculation of three-phonon scattering including thermal expansion in 
lower temperature phases (orthorhombic
phase and tetragonal phase) is computationally 
prohibitive since they have lower symmetry and more atoms in the primitive cell compared 
with the cubic phase. %
On the other hand, based on the experimental investigations, the coefficient 
of thermal expansion for CsPbBr$_3$ is 3.8 $\times$  10$^{-5}$ K$^{-1}$, 6.5 $\times$  10$^{-5}$ K$^{-1}$, 2.6 $\times$  10$^{-5}$ K$^{-1}$ for orthorhombic, tetragonal, and cubic phase, respectively~\cite{haeger2020thermal}.  %
The effect of the lattice constant on the thermal conductivity might be neglected in
a moderate temperature range. %
Therefore, in our calculation, we neglect the thermal expansion like previous works~\cite{simoncelli2019unified,lee2017ultralow,lanigan2021two}. %
Most recently, an effective one-body Hamiltonian that well represents the quasiparticle-peak have been developed. In this method, the thermal expansion in the calculation is included~\cite{tadano2022first}. %

\section{IV. CONCLUSIONS}
In summary, our study reveals the significant contributions of four-phonon 
scattering and the off-diagonal terms of the heat flux operators in calculating the thermal conductivity
in systems with harmonic phonons exhibiting imaginary frequencies and 
temperature renormalization. %

Specifically, our investigation of CsPbBr$_3$ halide perovskites in 
orthorhombic, tetragonal, and cubic phases yields the following key findings: %

(i) For materials with ultralow lattice thermal conductivity $\kappa_L$, the 
inclusion of high-order anharmonicity and off-diagonal terms bridge the gap between experimental observations and theoretical predictions; %

(ii) The strong coupling between high-frequency optical
phonons and overdamped acoustic phonons ($I_{\lambda\lambda_1}$), provides
insights into the intriguing phonon renormalization phenomena observed in 
strongly anharmonic systems as a function of temperature; %

(iii) Beyond the conventional phonon-phonon scattering perspective, phenomena 
such as electron-phonon coupling, polaron formation, and entropy in halide 
perovskites warrant further theoretical 
advancements~\cite{zhou2018electron, lanigan2021two}. %

Our study that presents an effective approach to understand the ultralow $\kappa_L$ observed in halide perovskites, might inspire further experimental investigations exploring materials with glass-like thermal 
conductivity. %

\section{ACKNOWLEDGMENTS}
We acknowledge the support from the National Natural Science Foundation of China 
(No.12104356 and No.52250191). 
Z.G. acknowledges the support of China Postdoctoral Science Foundation (No.2022M712552), %
the Opening Project of Shanghai Key Laboratory of Special Artificial Microstructure Materials 
and Technology (No.Ammt2022B-1), and the Fundamental Research Funds for the Central 
Universities. 
We also acknowledge the support by HPC Platform, Xi’an Jiaotong University. Y.X. acknowledges Portland State University Lab Setup Fund.
\bibliography{References.bib} 
 \end{document}